# Zero roll-off retinal MHz-OCT using an FDML-Laser

Julian Klee[1], Jan Philip Kolb[2], Christin Grill[1], Wolfgang Draxinger[1], Tom Pfeiffer[1] and Robert Huber[1]

[1]Institut für Biomedizinische Optik, Universität zu Lübeck, Lübeck, Peter-Monnik-Weg 4, 23562 Lübeck, Germany
[2]Medizinisches Laserzentrum Lübeck GmbH, Peter-Monnik-Weg 4, 23562 Lübeck, Germany

## ABSTRACT

Optical coherence tomography (OCT) applications like ultra-widefield and full eye-length imaging are of high interest for various diagnostic purposes. In swept-source OCT these techniques require a swept light source, which is coherent over the whole imaging depth. We present a zero roll-off 1060 nm Fourier Domain Mode Locked-Laser (FDML-Laser) for retinal OCT imaging at 1.7 MHz A-scan rate and first long-range imaging results with it. Several steps such as improved dispersion compensation and frequency regulation were performed and will be discussed. Besides virtually no loss in OCT signal over the maximum depth range of 4.6 mm and very good dynamic range was observed. Roll-off measurements show no decrease of the point-spread function (PSF), while maintaining a high dynamic range.

## 1. INTRODUCTION

Optical coherence tomography (OCT) [1] is an important imaging modality in ophthalmic diagnostics. Although most applications in this field do not require a long coherence length (which corresponds to the maximum ranging depth of the OCT) of more than 2 mm, some newer applications require coherence over larger imaging depths. These include full eye-length imaging [2] as well as ultra-widefield imaging. The latter requires a larger imaging range with low roll-off, due to the artificially introduced delay by the scanning geometry [3]. There are 1060 nm light sources for swept-source OCT that have the necessary coherence length for these applications, such as the MEMS-VECSEL [2] and other short cavity Lasers which feature coherence lengths in the order of centimeters or more [4]. For some ophthalmic OCT applications, multi-MHz A-scan rates are desirable and currently only Fourier Domain Mode Locked-Laser (FDML-Laser) can achieve this with sufficient performance [5]. However, long coherence lengths have only been demonstrated on a 1300 nm system [6] so far. Here we present first results of a 1060 nm system with zero roll-off over the imaging range. This has been achieved by further improving the FDML-Laser's dispersion characteristics and the introduction of frequency regulation to compensate for the remaining temperature variations of the insulated and temperature stabilized Laser.

## 2. METHODS

Figure 1 shows schematically the retinal FDML OCT setup. In contrast to our previous version [3], we included frequency regulation of the FDML sweep frequency, as well as an advanced triggering and scanner signal unit. Furthermore, the dispersion compensation has been improved compared to the last iteration.

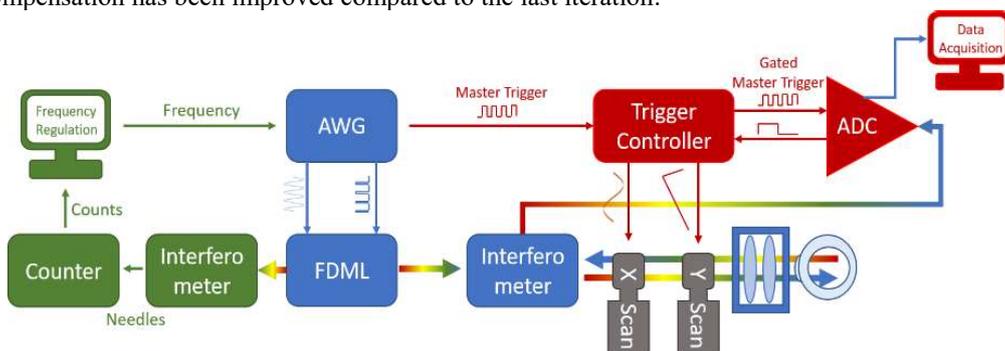

**Figure 1** An arbitrary waveform generator (AWG – Rigol DG1062Z) produces the signals necessary for driving the FDML-Laser as well as a master signal, which is used in the trigger controller to generate the scanner waveforms. It is also used for controlling the digitalization (analog digital converter - ADC) of the generated interference signal. Another interference signal is measured via an

1.6 GHz photodiode and analyzed by a frequency counter (Keysight 53230A) to set the correct frequency for the Laser. A specially designed new lens system is used to improve the quality of the image.

This was done by fine tuning the steps previously taken, which includes a new chirped Fiber Bragg Grating (cFBG) from Teraxion specified for 75 nm bandwidth. This compensates the strong dispersion caused by the 450 m long delay fiber. We used a combination of three different fiber types with different values of chromatic dispersion in order to compensate the remaining first and second order dispersion. This is required, because fibers and other components vary in dispersion, making an exact match of the cFBG impossible. Lastly a four-zone linear gradient cFBG heating element was used to fine tune the remaining third order dispersion to achieve *sweet-spot* operation [6].

Another improvement to the previous version was to incorporate the entire cavity and components like isolators into one tightly temperature-controlled fiber spool. This allows greater temperature stability in order to gain better frequency stability. Since a minute temperature variation change the cavity length and therefore the FDML frequency, a regulation of this frequency has been implemented in accordance with our previously demonstrated method for 1300 nm [6]. An interferometer with a delay of more than 2 cm between the two arms randomly produces frequencies within the detection bandwidth of the coupled photodiode (Wieserlabs BPD1GA), when the FDML-Laser does not operate in its *sweet-spot*. This becomes manifest in short transient features, which can be measured with a frequency counter (Keysight 53230A) and used in a live gradient descent optimization to adjust the frequency to maximize the *sweet-spot* of the Laser, where no such frequencies are created. In *sweet-spot* operation the FDML-Laser has drastically reduced amplitude and phase noise. The resulting FDML-Laser has a sweep frequency of 1.7 MHz at 75 nm bandwidth. The interference signal is measured with an 1.6 GHz dual balanced photodiode (Thorlabs PDB430C) at 2 GS/s digital conversion (AlazarTech ATS9397). Although this specific model supports 4 GS/s, the data packing for this is not yet implemented in our imaging setup.

The implementation of this frequency regulation as well as possible future phase-sensitive measurements necessitates a more complex trigger and scanner signal controller unit, which was implemented using an 128 MHz STMF4 ARM-based microcontroller and accompanying electronics board. This allows phase stable trigger gating when predefined conditions are met, also scanner signals can be produced, and a digital control can bias and amplify the signal. The unit also allows free frequency matching of the FDML frequency to scanners without influencing each other, which is done via frequency multiplication of the original master trigger. This is especially useful when using resonant scanners.

### 3. RESULTS AND DISCUSSION

Figure 2 displays the measured point-spread functions (PSF) at incremental delays over the whole imaging range.

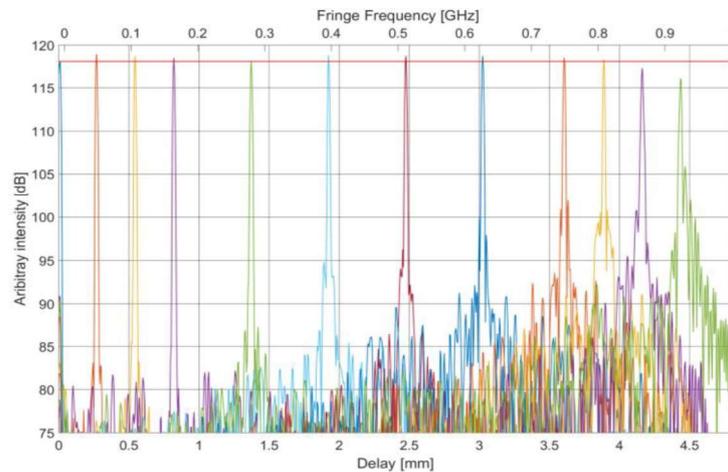

**Figure 2** Measurement of the PSFs over the whole imaging range. The red line indicates 1 dB loss from the maximum signal.

Virtually no roll-off in intensity of the PSF can be observed as indicated by the 1 dB line. The roll-off in the last PSFs is due to bandwidth limitations of the 2 GS/s analog to digital converter (ADC). In the non-bandwidth limited region, a minimum dynamic range (noise floor versus PSF peak) of 27 dB and a maximum of 44 dB can be observed. In the next step, we used the previously described imaging setup to perform OCT imaging on a healthy 31-year old volunteer, as seen in Figure 3. The retina is placed at different depth positions to evaluate signal loss and image degradation over depth. The results over a depth range of 3.75 mm are consistent with the roll-off measurements. These preliminary results are encouraging, because the fact that the signal is independent of the imaging depth, makes this system the ideal candidate for ultra-widefield retinal OCT systems.

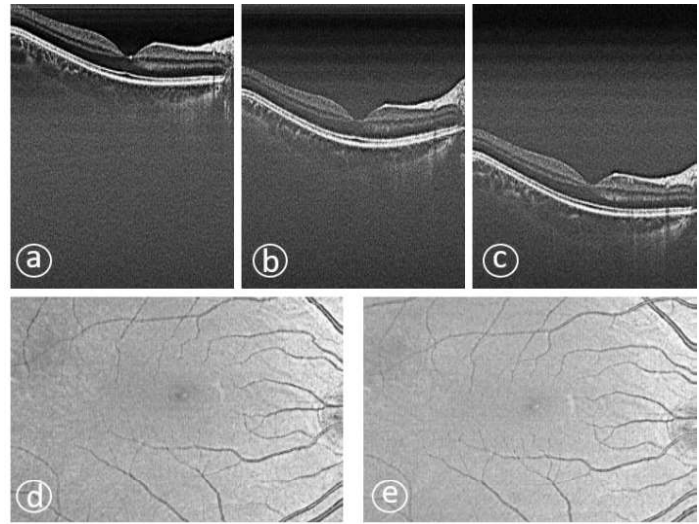

**Figure 3** Three B-frames (six times averaged) of the same volunteer (a-c) at different positions within the z-ranging depth of the OCT system as well as EnFace projections at two different depths (a and c) can be compared. Postprocessing for all images (a-c and d-e) was done with the same parameters.

## 4. CONCLUSION AND OUTLOOK

We demonstrated the first real roll-off data and long-range OCT imaging experiments with an FDML-Laser operating around 1060 nm wavelength in the *sweet-spot* operating regime. We observe very good coherence and sensitivity roll-off performance. To achieve this zero roll-off in FDML-Laser based swept-source retinal OCT a frequency regulation was implemented, which controls the sweep frequency. This required a sophisticated electronic synchronization approach in order to achieve the required accuracy. It has been shown that a high dynamic range in the image can be achieved using the described methods. In future research the phase stability of the FDML-Laser will be characterized and used for phase stable measurements. Ultra-widefield imaging will be performed to further evaluate the performance and make use of the long coherence.

## 5. ACKNOWLEDGEMENTS

This project was funded by the European Union project ENCOMOLE-2i (Horizon 2020, ERC CoG no. 646669), the German Research Foundation (DFG project HU1006/6 and EXC 306/2; DFG EXC 2167), the German Federal Ministry of Education and Research (BMBF no. 13GW0227B: "Neuro-OCT").